\documentclass[showpacs,prb,superscriptaddress,twocolumn,floatfix,reprint]{revtex4-1}
\usepackage[T1]{fontenc}
\usepackage[latin9]{inputenc}
\usepackage{textcomp}
\usepackage{amstext}
\usepackage[Symbol]{upgreek}
\usepackage{graphicx}
\usepackage{amsmath}
\usepackage{booktabs}
\usepackage{hyperref}
\usepackage{dcolumn}
\usepackage{enumerate}
\usepackage{latexsym,bm}
\makeatletter

\begin{document}
\preprint{manuscript}

\title{Roles of oxygen vacancy on ferromagnetism in Ni doped In$_2$O$_3$: A hybrid functional study}

\author{V. Wang}
\thanks{Corresponding author at: Department of Applied Physics, Xi'an University of Technology, No.58, Yanxiang Road, Xi'an 710054, China, Tel./Fax: +86-29-8206-6357/6359 \\E-mail address: wangvei@me.com (V. Wang).}
\affiliation{Department of Applied Physics, Xi'an University of Technology, Xi'an 710054, China}  

\author{C.-Y. You}
\affiliation{School of Materials Science and Engineering, Xi'an University of Technology, Xi'an 710048, China}  

\author{H.-P. He}
\affiliation{Department of Geological Engineering, Lanzhou Resources \& Environment Voc-Tech College,
Lanzhou 730021, China}

\author{D.-M. Ma}
\affiliation{Department of Applied Physics, Xi'an University of Technology, Xi'an 710054, China}

\author{H. Mizuseki}
\affiliation{Institute for Materials Research, Tohoku University,
Sendai 980-8577, Japan}

\author{Y. Kawazoe}
\affiliation{Institute for Materials Research, Tohoku University,
Sendai 980-8577, Japan}

\date{\today}
\begin{abstract}                                   
\begin{center}
\textbf{ABSTRACT}
\par\end{center}
The roles of oxygen vacancies on the electronic and magnetic properties of Ni doped In$_2$O$_3$ have been studied by first-principles calculations based on hybrid functional theory. 
Our results predict that the Ni-doped In$_2$O$_3$ system displays a ferromagnetic semiconducting character.
However, the presence of oxygen vacancies results in antiferromagnetic coupling between the neighboring Ni pair bridged by an oxygen vacancy. The antiferromagnetic coupling is found to arise from the predominant role of superexchange due to the strong Ni 3\emph{d}-O 2\emph{p} hybridization. Consequently, the oxygen vacancies play a key role in the lower saturation magnetization of Ni:In$_2$O$_3$ polycrystalline sample, as observed in experiments.
\end{abstract}
 


\maketitle

\section{introduction}                             

\noindent Dilute magnetic semiconductors (DMSs) have been extensively
investigated due to their potential applications for spintronic devices,
\cite{Ohno1998281,Ohno1999402,Fiederling,dietl,Zunger2010,Sato2010} which are obtained by doping conventional semiconductor materials with transition metal impurities causing ferromagnetic (FM) ordering. A large number of experimental studies indicate that the DMSs with \emph{T$_{\text{c}}$} higher than 300 K
are mainly concentrated in transition-metal doped wide band-gap semiconductors, such as
various oxides \cite{Ueda,Matsumoto,Han,Shinde,Ogale,Coey,Sluiter,Philip}
and nitrides. \cite{Teraguchi,Wu,Liu,Kumar}

In$_2$O$_3$ is a transparent semiconductor with a wide direct band-gap of around 2.93 eV.\cite{Walsh2008,King2009}  It is an excellent material which can integrate electronic, magnetic, and photonic properties in new generation devices. Recent experimental findings show that V\cite{Gupta2007}, Co\cite{Hong2006}, Fe\cite{He2005,Yu2006}, Cu\cite{Yoo2005}, Cr\cite{Kharel2007,Philip2006} and Ni\cite{Hong2005} doped In$_2$O$_3$ systems have attracted much attention because of the observation of room-temperature ferromagnetism. Meanwhile, several controversial results on the magnetic properties of Ni-doped In$_2$O$_3$ have been reported: Hong \emph{et al}. fabricated well-crystallized Ni:In$_2$O$_3$ films by using laser ablation and they claimed that these films show room temperature ferromagnetism with magnetic moments of 0.7 $µ_B$/Ni.\cite{Hong2005} Peleckis \emph{et al}.\cite{Peleckis2006} pointed that the Ni:In$_2$O$_3$ polycrystalline samples obtained with solid state synthesis show a typical ferromagnetism character, however, with a very small average magnetic moment of 0.03-0.06 $µ_B$/Ni at 300 K. They speculated that the non-negligible concentration of oxygen vacancies (V$_\text{O}$) existing in these samples which prepared under argon atmosphere might be responsible for such magnetic phenomena. These findings suggest that the magnetic behavior of Ni doped In$_2$O$_3$ is strongly dependent on the preparation methods. Like other metal-oxide semiconductors, such as ZnO and SnO$_2$, several native defects can form during the growth of In$_2$O$_3$. Among these, the most common defects are oxygen vacancies. \cite{Wit1975,Wit1977,Wit1977a,Lany2007,Agoston2009} 
A considerable amount of experimental data have revealed that oxygen vacancy play a key role in the ferromagnetism of transition metal doped ZnO, SnO$_2$ and In$_2$O$_3$.\cite{Kundaliya2004,Hsu2006,Ramachandran2006,Hong2005a,Archer2005,Coey2004} Previous theoretical studies based on traditional density functional theory (DFT), 
or DFT plus Hubbard U focused on the electronic structure of transition-metal doped In$_2$O$_3$ without native defects. \cite{Raebiger2009,Huang2009}
There are few investigations about the roles of oxygen vacancies on magnetism in Ni doped In$_2$O$_3$. In addition, it is well known that traditional DFT can not well describe the localized character of \emph{d}-band electrons. The physics of localized 3\emph{d} states can be partially described using a DFT+U scheme. However, accurate electronic properties are not completely recovered, such as the band gap, especially in oxide semiconductors. Even in the Ge, the half metallicity of this compound is lost within DFT+U scheme.\cite{Stroppa2011}

Recently, a hybrid density functional approach, which admixes the non-local Hartree-Fock exchange into traditional local LDA or semilocal GGA exchange-correlation functionals, has been reported to provide an improved description of the electronic structure for a variety of extended solid-state systems.\cite{Adamo1999,Heyd2003,Paier2006} From the above discussions, it is clear that a detailed investigation of electronic structure and magnetic interactions is necessary, taking into account the effects of V$_\text{O}$ on electronic and magnetic properties in Ni doped In$_2$O$_3$. The remainder of this paper is organized as follows. In Sec. II, the details of the computation are described. Sec. III presents our calculated results with respect to the electronic and magnetic properties of Ni doped In$_2$O$_3$ without and with V$_\text{O}$. Finally, a short summary is given in Sec. IV.

\section{methods}                                  

Our total energy and electronic structure calculations were based on the hybrid density functional as proposed by Heyd, Scuseria, and Ernzerhof (HSE) \cite{Heyd2003} and the projector augmented wave potentials \cite{paw} as
implemented in the VASP code.\cite{Kresse1994,Kresse1999} 
To reproduce the experimental band gap of $\text{In}_2\text{O}_3$, a screening parameter of 0.2 {\AA}$^{-1}$ and a mixed proportion
of 32\% Hartree-Fock exchange with 68\% GGA of Perdew, Burke and Ernzerhof (PBE) \cite{Perdew1996} exchange were used for the HSE06 functional.\cite{Krukau2006} The semi-core In \emph{d}
electrons were treated as core electrons. Varley \emph{et al.} have claimed that such treatment leads to the deviation in the formation energies of defects less than 0.1 eV with respect to the case of the In \emph{d} electrons treated as valence states.\cite{Varley2011} The valence electrons
configuration for nickel, indium as well as oxygen are [Ni] $3\emph{d}^84\emph{p}^2$, [In] $5\emph{s}^25\emph{p}^1$  and [O] $2\emph{s}^22\emph{p}^4$ respectively.

\begin{figure}[htbp]
\centering
\includegraphics[scale=0.44]{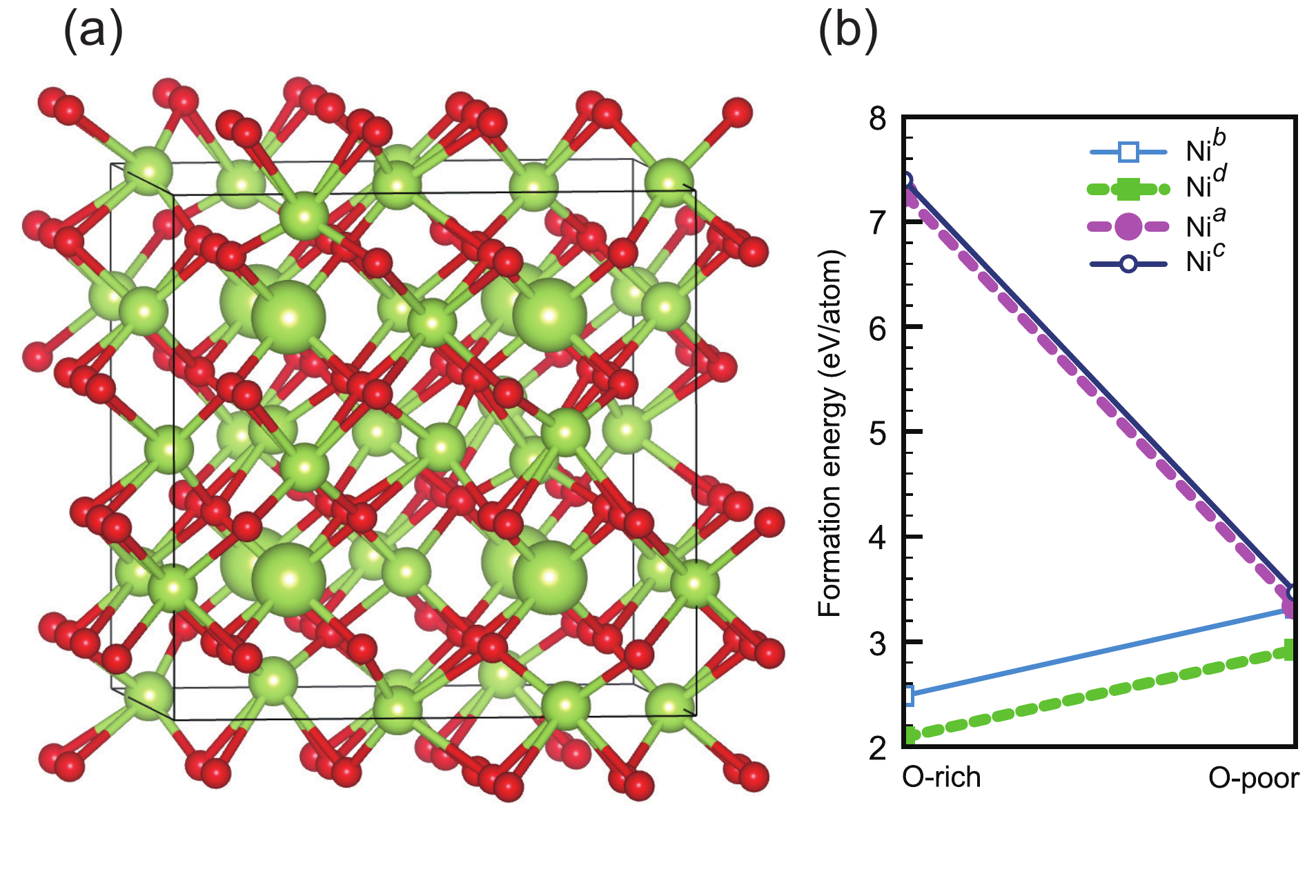}
\caption{\label{structure}(Color online) (a) Supercell used to model Ni doped In$_2$O$_3$ bixbyite structure. The In$^b$, In$^d$ and O atoms are represented with big green, small green and red balls respectively. (b) Formation energies of substitutional (Ni$^b$ and Ni$^d$) and interstitial (Ni$^a$ and Ni$^c$) Ni in In$_2$O$_3$.}
\end{figure}

In$_2$O$_3$ crystallizes in a cubic bixbyite type structure which has 80 atoms (32 In and 48 O atoms) in its unit cell.\cite{In2O3}
According to the Wyckoff's notation, it has two inequivalent In sites: there are eight In atoms occupying the \emph{b} sites and 24 In atoms occupying the \emph{d} sites, labeled by In$^b$ and In$^d$ respectively. Both In$^b$ and In$^d$ are surrounded by six O atoms. On the other hand, 48 oxygen atoms occupy the \emph{e} sites which are four-fold coordinated surrounded by three In$^d$ and an In$^b$ atoms. For the calculations of Ni doped systems, we employed a supercell containing 80 atoms, i.e., one unit cell of bixbyite In$_2$O$_3$, as shown in Fig. \ref{structure} (a). The calculated equilibrium lattice constant of In$_2$O$_3$ is 10.13 \AA, which agrees well the experiment value of 10.12 \AA.\cite{Gonzalez2004}  
A 2$\times$2$\times$2 mesh within Monkhorst-Pack scheme\cite{Monkhorst1976} and Gaussian smearing of 0.05 eV was applied to the Brillouin-zone integrations in total-energy calculations. The wave functions were expanded by plane waves up to
a cutoff energy of 300 eV. The internal coordinates in the supercells consisting of Ni dopants were relaxed to reduce the residual force to less than 0.02 eV$\cdot$\AA$^{1-}$, with the lattice constants fixed at the optimized values for the perfect crystal. 

In this study, two ways are used for introducing a Ni dopant into the
supercell. One is to replace an In atom with a Ni one, and the other is to introduce an interstitial Ni atom into the supercell. Ni dopants can substitute indium atoms on In$^b$ sites or In$^d$ sites, represented by Ni$^b$ and Ni$^d$. There are two possible interstitial Ni sites: \cite{Warschkow2006,Palandage2010} The Ni$^c$-site (Wyckoff 16\emph{c}) is tetrahedrally coordinated by four cations (In) with Wyckoff coordinates (\emph{x}, \emph{x}, \emph{x}), where \emph{x} = 0.116 is in fractional-coordinated units. The Ni$^a$-site [Wyckoff coordinates (0, 0, 0)] depicts an octahedral position in the sub-lattice of six \emph{d}-site indium atoms. 

\section{Results and discussion}                   

We start with examining the formation energy of a single Ni atom occupying a substitutional or interstitial site, corresponding to a Ni-doping concentration of about 4.2\%. The formation energies of Ni dopants, also depend on the chemical potentials of Ni atom and the host elements (In and O), i.e., on the relative abundance of these elements in the growth environment. In our current study, the chemical potentials $\mu_\text{O}$, $\mu_\text{In}$ and $\mu_\text{Ni}$ are referenced to the calculated energy of per atom in an isolated oxygen molecule, tetragonal phase of indium and face-center-cubic phase of nickel, respectively. They thereby are subject to upper bounds $\mu_\text{O}$<=0 (O-rich limit) and $\mu_\text{In}$<=0 (O-poor limit) and must satisfy the stability condition of In$_2$O$_3$: 2$\mu_\text{In}$+3$\mu_\text{O}$=$\Delta$\emph{H}(In$_2$O$_3$), where $\Delta$\emph{H}(In$_2$O$_3$)=-9.53 eV is the formation energy of In$_2$O$_3$. To avoid the formation of NiO as a secondary phase, the chemical potential $\mu_\text{Ni}$ is limited by $\mu_\text{O}$+$\mu_\text{Ni}$<=$\Delta$\emph{H}(NiO), where $\Delta$\emph{H}(NiO)=-3.93 eV is the formation energy of NiO. 

Figure \ref{structure} (b) presents the calculated formation energies of substitutional Ni and interstitial Ni under both O-rich and O-poor conditions in In$_2$O$_3$. Our results show that 
the In$^d$ site is the most energetically favorable one for Ni occupying under both sets of growing conditions. Even at the oxygen-poor limit, the Ni$^d$ is still more stable than the Ni$^a$ (the most energetically favored interstitial site) by 0.40 eV. The small formation energy difference between Ni$^a$ and Ni$^c$ indicates that no interstitial site is favorable.
As for the local lattice relaxation around the Ni dopants, similar to what was done in our previous study,\cite{Wang2012} we define elastic deformation energy ($\Delta$E$_{ed}$) to evaluate the scale of distortion accompanied by Ni incorporated into In$_2$O$_3$ host. The calculated values of $\Delta$E$_{ed}$ induced by Ni$^d$, Ni$^b$ and Ni$^a$ (Ni$^c$) are 1.53 eV, 0.70 eV and 0.16 eV respectively. More specifically, the four of the six nearest neighbor O atoms relax inwards by about 10\% (0.19 \AA) to give a Ni$^d$-O bond length of 1.95 \AA; while the rest two nearest neighbor O atoms slightly relax outward by about 1\% (0.03 \AA). The Ni$^b$ defect caused a similar inward relaxation for all the six nearest neighbors, but to a lesser extent of 6\% (0.14 \AA), giving a Ni$^b$-O bond length of 2.05 \AA. In contrast, the local lattice distortions around the interstitial Ni$^a$ and Ni$^c$ can be ignored when compared to the substitutional Ni defects. This suggests that the interstitial sites are sufficient enough to accommodate Ni atom.

It is noteworthy that the competition between the Ni-O bonding interaction and the elastic deformation introduced by Ni determines the stability of various type of Ni defects in In$_2$O$_3$. Bader charge analyses \cite{Henkelman2006} estimate that Ni$^d$, Ni$^b$ and Ni$^a$ (Ni$^c$) denote about 1.6\emph{e}, 1.5\emph{e}, and 1.1\emph{e} to their neighboring O atoms respectively, indicating the typical ionic character of Ni-O bond.
This implies that the more charges transfer from Ni to O atoms, the stronger the bonding strength between them will be. Consequently, the strongest bonding interaction occurs in Ni$^d$-O bond, and Ni$^b$-O bond, then followed by Ni$^a$ (Ni$^c$) -O bond. The former in turn leads to the remarkable local lattice distortion. 
Note that the Ni$^d$ is more stable by at least 0.39 eV as compared with the other types of Ni dopant. We therefore here only consider Ni ions incorporated substitutionally on the In$^d$ sites in the following studies.

To determine whether Ni dopants attract or repel each other, we calculated the total energies of two configurations in which two Ni atoms are placed either in the closest (3.4 \AA) or furthest (8.5 \AA) separation distance (Ni concentration: 8.4\%). We labeled these two configurations as 2Ni$^d$-near and 2Ni$^d$-far. In addition, as well be discussed later, an oxygen vacancy is introduced by removing one oxygen atom adjacent to or far away Ni dopants. For all doped configurations, both FM and antiferromagnetic (AFM) orderings were calculated in order to find the magnetic ground state. We defined the FM stabilization energy as $\Delta$E$_\text{FM}$ = 
E$_\text{FM}$-E$_\text{AFM}$, where E$_\text{FM}$  and E$_\text{AFM}$ are the total energies of the supercell with FM and AFM orderings, respectively. A negative value of $\Delta$E$_\text{FM}$ means that the FM states is favorable. Our main calculated results for the considered configurations are summarized in TABLE \ref{table1}, involving the formation energies (under the corresponding most stable magnetic ordering), FM stabilization energies and magnetic moments. In the absence of oxygen vacancy (V$_\text{O}$), one can see that the formation energy difference of Ni atom in the 2Ni$^d$-near and 2Ni$^d$-far configurations is only 0.07 eV/Ni, suggesting that Ni dopants have no significant trend to cluster. In other words, the distribution of Ni could be considered to be uniform. The FM phase is found to be energetically favored more than AFM one in two considered cases. These findings are in good agreement with the recent experiments of Hong \emph{et al.} who found that Ni impurities distribute largely uniform in the Ni:In$_2$O$_3$ samples.\cite{Hong2005} It should be pointed that the separation distance between two Ni ions is so far that the FM magnetic coupling in the 2Ni$^d$-far configuration becomes rather weak (around 3 meV/cell).

\begin{figure}[htbp]
\centering
\includegraphics[scale=0.4]{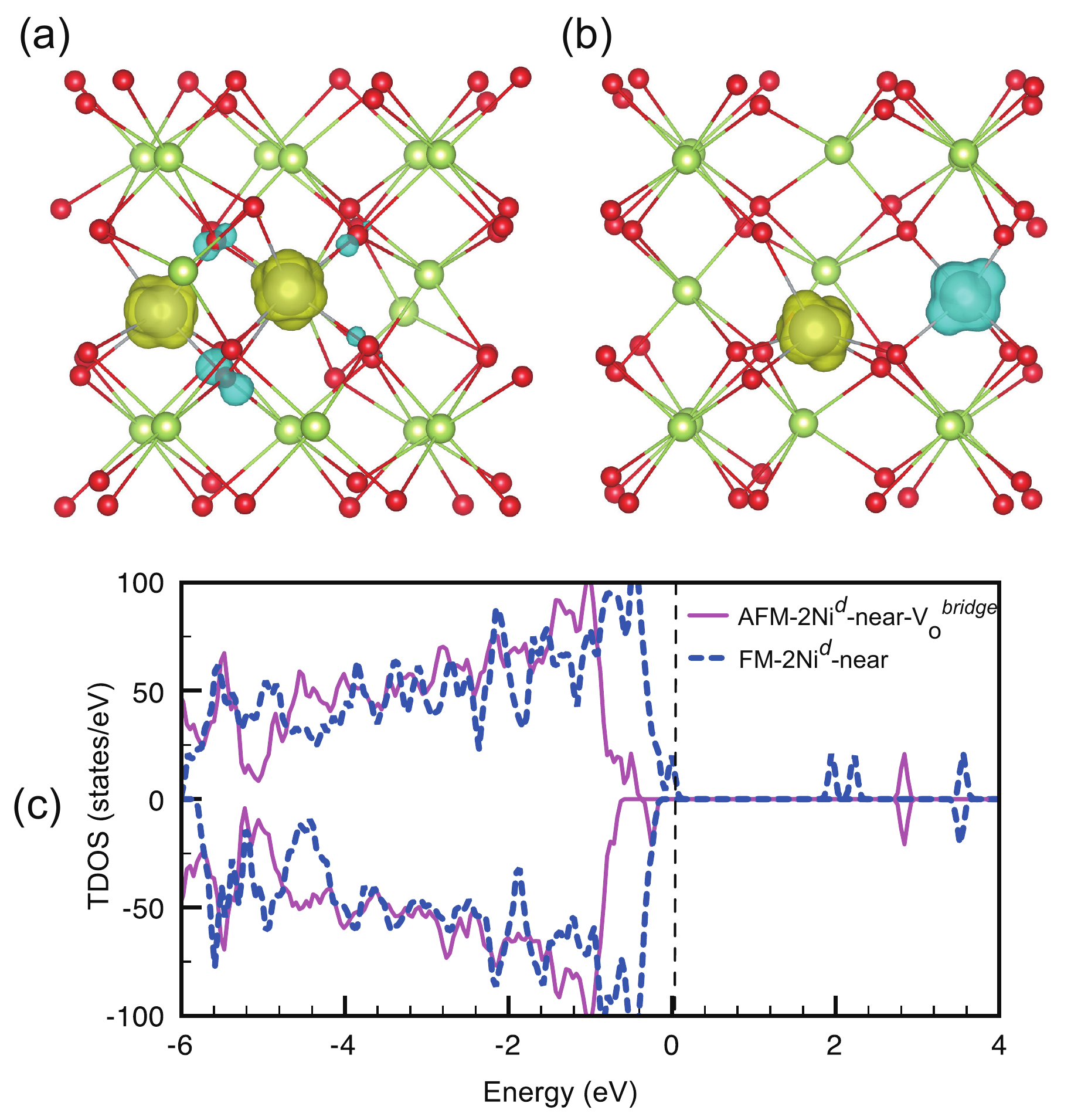}
\caption{\label{spinden}(Color online) Spatial spin-density distribution of magnetic ground states for (a) 2Ni$^d$-near and (b) 2Ni$^d$-near-V$_\text{O}^{bridge}$ configurations. Yellow and blue isosurfaces corresponding to spin-up and spin-down regions (isosurface: 3$\times$10$^{-2}$ electrons/bohr$^{3}$). (c) Total DOS of these two configurations. Positive (negative) values refer to spin-up (-down) component. The Fermi energy is set to zero.}
\end{figure}

Next we introduced a V$_\text{O}$ into the 2Ni$^d$-near configuration. We found that the total energy of the configuration with a  V$_\text{O}$ on the bridged oxygen (O$^{bridge}$) site of the two Ni neighbors (denoted as 2Ni$^d$-near-V$_\text{O}^{bridge}$) is energetically more stable by around 1.39 eV than that of the configuration with a furthest separation distance (8.5 \AA) between V$_\text{O}$ and Ni pair.
This implies that V$_\text{O}$ is attracted by its Ni neighbors to form a complex. Also, as presented in TABLE \ref{table1}, the average formation energy of Ni significantly reduces to 0.89 (1.64) eV/Ni under the O-poor (O-rich) conditions due to the presence of neighboring V$_\text{O}^{bridge}$. This indicates that V$_\text{O}$ can enhance the solubility of Ni in In$_2$O$_3$ host, similar findings were observed in Co doped SnO$_2$.\cite{Fitzgerald2006} However, a dramatic change in the type of favorable magnetic ordering in 2Ni$^d$-near-V$_\text{O}^{bridge}$ is very surprising, where the $\Delta$E$_\text{FM}$ is a positive value. This implies an energetic AFM favorable ground state. 
The calculated spatial spin-density distributions of Ni doped In$_2$O$_3$ systems without and with V$_\text{O}^{bridge}$ are plotted in Fig. \ref{spinden} (a) and (b) respectively. It is found that the total magnetic moment are mainly derived from Ni 3\emph{d} states. Each Ni atom contributes about 1.3 $\mu_B$/Ni to the system (see TABLE \ref{table1}); while the nearest-neighboring O atoms give a small negative contribution, less than 0.3 $\mu_B$ per O atom. As a result, the host semiconductor with a Ni concentration of 8.4\% carries a integer total magnetic moment of 2 $\mu_B$/cell. Since the introduction of V$_\text{O}^{bridge}$ leads a small fraction of valence charges back to the Ni 3\emph{d} states, the local magnetic moment of Ni is slightly enhanced, reaching up to 1.8 $\mu_B$/Ni. However, the total magnetic moment of 2Ni$^d$-near-V$_\text{O}^{bridge}$ configuration decreases to zero as it has a AFM magnetic ground state. Further studies show that AFM ordering diminishes rapidly when the V$_\text{O}$ departs from Ni dopants, suggesting the localized character of interaction among V$_\text{O}$ and Ni dopants.

\begin{figure}[htbp]
\centering
\includegraphics[scale=0.45]{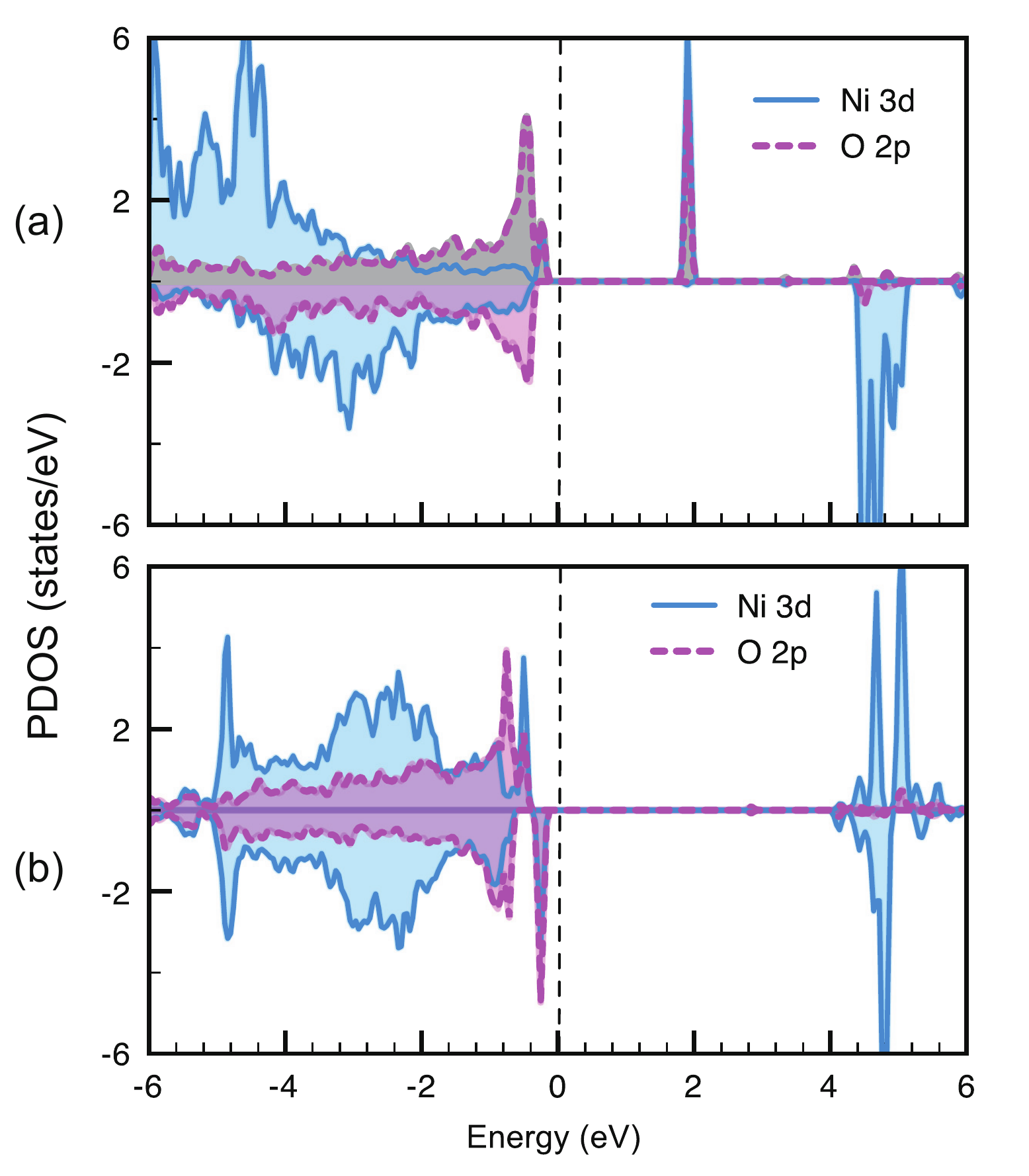}
\caption{\label{pdos}(Color online) Projected DOSs (PDOSs) of Ni and O atoms. (a) and (b) for AFM ordering of system without  and with V$_\text{O}$ respectively. Positive (negative) values refer to spin-up (-down) component. The Fermi energy is set to zero.}
\end{figure}

From the total density of states (TDOS) displayed in Fig. \ref{spinden} (c), one can find that the Ni doped systems without and with V$_\text{O}^{bridge}$ maintain the semiconducting character of the host material. Due to the lack of itinerant carriers, the ferromagnetism of insulating In$_2$O$_3$ with Ni doping is likely arise from a percolation of bound magnetic polarons,\cite{Wolff1996,Kaminski2002} instead of the carrier-mediated magnetism proposed for conducing DMS. In view of the fact that the magnetic coupling between Ni dopants switches to AFM after the introduction of V$_\text{O}^{bridge}$, the question is whether a new exchange mechanism emerges in this process. To further shed light on this issue, we plotted the calculated projected density of states (PDOS) of AFM orderings for 2Ni$^d$-near and 2Ni$^d$-near-V$_\text{O}^{bridge}$ configurations in Fig. \ref{pdos} (a) and (b). In the perfect Ni doped system, the main peak of Ni 3\emph{d} states is found to be lying 2-6 eV below the valence band maximum (VBM) while the main proportion of O 2\emph{p} states is distributed near the VBM. Thus, a small hybridization between Ni 3\emph{d} and O 2\emph{p} states can only form near the VBM. The calculated O 2\emph{p} states in the host material extends around -5.6 eV to -0.3 eV (not shown), in good consistent with the experimental values of O 2\emph{p} bandwidth observed in In$_2$O$_3$.\cite{Piper2009,Janowitz2011,Green2012,McLeod2012} As shown in Fig. \ref{pdos}, the O 2\emph{p} states become more delocalized as we placed a Ni atom at the neighboring cation site. 
For the 2Ni$^d$-near configuration, the two Ni atoms are bridged by two O$^{bridge}$ atoms.
When we introduced a V$_\text{O}^{bridge}$ into this configuration by removing one of these two O$^{bridge}$ atoms, we found that the Ni 3\emph{d} states shift up in energy and significantly overlap with the 2\emph{p} states of the remaining O$^{bridge}$ in the range of -6-0 eV. 
It is clearly shown that a strong hybridization between Ni 3\emph{d} and O 2\emph{p} is observed from the PDOS of AFM ordering for 2Ni$^d$-near-V$_\text{O}^{bridge}$ as displayed in Fig. \ref{pdos} (b). Further calculations show that the FM and AFM couplings become degenerate in energy if we remove all two V$_\text{O}^{bridge}$ atoms, confirming the key role played by O$^{bridge}$ in the magnetic interaction between two neighboring Ni atoms. Thus, the AFM coupling between Ni atoms might be the result of the fact that the superexchange is predominant over in the magnetic mechanisms. In addition, we also found that the O 2\emph{p} bandwidth does not change at higher concentration of V$_\text{O}$.

Based our calculated results, we give an explanation to the experimental controversial results involved the magnetization of Ni-doped In$_2$O$_3$ system, where the key role is V$_\text{O}$.
In the absence of V$_\text{O}$, a FM magnetic ground state with a calculated moment of 1.0 $\mu_\text{B}$/Ni is predicted in the perfect Ni-doped In$_2$O$_3$.
It should be pointed that the calculated magnetic moment of Ni is slightly larger than the experimental value (0.7 $µ_B$/Ni) reported in the well-crystallized samples.\cite{Hong2005} We attribute this difference to the unavoidable formation of V$_\text{O}$ in these samples even under the well-controlled growth conditions. With the existence of V$_\text{O}$, the Ni atoms are attracted to occupy the  neighboring In sites of V$_\text{O}$. The V$_\text{O}$ acts as a FM coupling killer to make these neighboring Ni dopants show AFM coupling. As a consequence, the strength of FM coupling and the effective magnetic moment of the doped system are lessened. 
It is believed that the higher concentration of V$_\text{O}$ would lead the lower portion of Ni dopants to contribute to the effective magnetic moment of the doped system. Wit \emph{et al.} \cite{Wit1975,Wit1977,Wit1977a} have reported that the concentration of V$_\text{O}$ in the In$_2$O$_3$ crystals can reach up to 1\% even under equilibrium growth conditions. Such results provide direct evidence that the significant concentration of V$_\text{O}$ plays an important role in explanation of the very small average local magnetic moment of Ni atom (only 0.03-0.06 $µ_B$/Ni at 300 K) observed in Ni:In$_2$O$_3$ polycrystalline samples. 

\begin{table*}[htbp]
\centering
\caption{\label{table1} Formation energy ($\Delta$E$_\text{F}$, in eV/Ni), FM stabilization energy ($\Delta$E$_\text{FM}$, in meV/cell), total (M$_\text{total}$, in $\mu_\text{B}$/cell) and local magnetic moments of Ni (M$_\text{Ni}$, in $\mu_\text{B}$/Ni) in Ni doped systems with and without V$_\text{O}$.}
\begin{ruledtabular}
\begin{tabular}{cccccccc}
&\multicolumn{3}{c}{}
&\multicolumn{2}{c}{FM} 
&\multicolumn{2}{c}{AFM}\\
Configuration &$\Delta$E$_\text{F}$ (O-rich) & $\Delta$E$_\text{F}$ (O-poor) & $\Delta$E$_\text{FM}$ & M$_\text{total}$ & M$_\text{Ni}$ & M$_\text{total}$ &M$_\text{Ni}$\\
\hline
2Ni$^d$-near & 1.93 & 2.77& -16& 2.0  & 1.36 (1.27)& 0.0 & 1.33 (-1.24)\\
2Ni$^d$-far & 2.00 & 2.84 & -3 & 2.0 & 1.30 (1.30) & 0.0 & 1.31 (-1.30)\\
2Ni$^d$-near-V$_\text{O}^{bridge}$ & 1.64 & 0.89& 15 & 4.0  & 1.77 (1.78) & 0.0 & 1.76 (-1.78)\\
\end{tabular}
\end{ruledtabular}
\end{table*}

\section{summary}                                  

In summary, the effects of V$_\text{O}$ on the ferromagnetism of Ni doped In$_2$O$_3$ systems were systematically studied based on the hybrid density electronic structure calculations. Our calculated results demonstrate that Ni dopants distribute uniformly with FM coupling in In$_2$O$_3$ host. With the existence of V$_\text{O}$, the Ni dopants are attracted to occupy the neighboring In sites of V$_\text{O}^{bridge}$ to form a complex. 
As a result, the rest one O atom bridged two Ni neighbors acts to a superexchange mediator by causing an indirect AFM coupling between these two Ni atoms. Thus, the high concentration of V$_\text{O}$ would be responsible for the smaller effective magnetic moments of Ni dopants experimentally observed in In$_2$O$_3$ polycrystalline samples.
  
\begin{acknowledgments}
Wang acknowledges the support of the Natural National Science Foundation of Shaanxi Province (Grant No. 2013JQ1021). You acknowledges the support of the NSFC (No. 51171148). Ma acknowledges the support of the Scientific Research Program Funded by Shaanxi Provincial Education Department, China (Program No. 2010JK743). The calculations were performed on the HITACHI SR16000 supercomputer at the Institute for Materials Research of Tohoku University, Japan.
\end{acknowledgments}

\nocite{*}
\bibliographystyle{aipnum4-1}

\end{document}